%% file: yoda2.tex
\pdfoutput=1

\PassOptionsToPackage{x11names,svgnames}{xcolor}

\documentclass{SciPost}
\input{preamble}

\hypersetup{hidelinks,
  pdfauthor={Andy Buckley, Matthew Filipovich, Christian Gutschow, Nick Rozinsky, Simon Thor, Yoran Yeh, Jamie Yellen}
  pdftitle={Consistent, multidimensional differential histogramming and summary statistics with YODA 2}
}

\begin{document}
\pagestyle{SPstyle}

%% Title
\begin{center}
  \color{scipostdeepblue}
  \Large
  %\textbf{\yoda[2]: performant and expressive binning\\ and summary statistics in \cxx and \python}
  \textbf{Consistent, multidimensional differential histogramming\\ and summary statistics with \yoda[2]}
\end{center}

%% Author list: use initials + surname format.
% Mark the corresponding author with a superscript *.
\begin{center}
  A.~Buckley$^1$,
  L.~Corpe$^2$,
  M.~Filipovich$^3$,
  C.~G\"utschow$^{4,5,*}$,\\
  N.~Rozinsky$^1$,
  S.~Thor$^6$,
  Y.~Yeh$^5$,
  J.~Yellen$^1$
\end{center}

\begin{center}
  \itshape
  $^1$ School of Physics \& Astronomy, University of Glasgow,\\ University~Place, G12~8QQ, Glasgow, UK\\
  $^2$ LPC, Universit\'e Clermont Auvergne, CNRS/IN2P3, Clermont-Ferrand, France\\
  $^3$ Department of Physics, University of Oxford, Clarendon Laboratory,\\ Parks Road, Oxford, OX1 3PU, UK\\
  $^4$ Centre for Advanced Research Computing, University College London,\\ Gower Street, London, WC1E~6BT, UK\\
  $^5$ Department of Physics \& Astronomy, University College London,\\ Gower~Street, WC1E~6BT, London, UK\\
  $^6$ KTH Royal Institute of Technology, SE-100 44 Stockholm, Sweden\\
\end{center}

\begin{center}
\today
\end{center}

%% Disable when final
% \linenumbers

\section*{Abstract}
Histogramming is often taken for granted, but the power and
compactness of partially aggregated, multidimensional summary
statistics, and their fundamental connection to differential and
integral calculus make them formidable statistical objects, especially
when very large data volumes are involved. But expressing these
concepts robustly and efficiently in high-dimensional parameter spaces
and for large data samples is a highly non-trivial challenge -- doubly
so if the resulting library is to remain usable by scientists as
opposed to software engineers. In this paper we summarise the core
principles required for consistent generalised histogramming, and use
them to motivate the design principles and implementation mechanics of
the re-engineered \yoda histogramming library, a key component of
physics data--model comparison and statistical interpretation in
collider physics.

%% Guideline: if your paper is longer that 6 pages, include a TOC
\clearpage
%% \vspace{20pt}
%% \noindent\rule{\textwidth}{1pt}
\tableofcontents
\thispagestyle{fancy}
%% \noindent\rule{\textwidth}{1pt}
%% \vspace{10pt}
\clearpage

%% Contents in separate files for each section
\input{sec-intro}
\input{sec-stats}
\input{sec-design}
\input{sec-yoda1}
\input{sec-impl}
\input{sec-formats}
\input{sec-plot}
\input{sec-concl}

\section{Acknowledgements}
The authors thank the Marie Sklodowska-Curie Innovative Training
Network MCnetITN3 (grant agreement no. 722104) for funding and
providing the scope for discussion and collaboration toward this
work. AB and CG acknowledge funding via the STFC experimental
Consolidated Grants programme (grant numbers %
ST/S000887/1 \& %< 2019-23 https://gtr.ukri.org/projects?ref=ST%2FS000887%2F1
ST/W000520/1 %< 2022-25 https://gtr.ukri.org/projects?ref=ST%2FW000520%2F1
and ST/S000666/1) \& %< 2019-23 https://gtr.ukri.org/projects?ref=ST%2FS000666%2F1
ST/W00058X/1), %< 2022-25 https://gtr.ukri.org/projects?ref=ST%2FW00058X%2F1
and the SWIFT-HEP project (grant numbers
ST/V002562/1 % Sheff/main: https://gtr.ukri.org/projects?ref=ST%2FV002562%2F1
and ST/V002627/1). % UCL: https://gtr.ukri.org/projects?ref=ST%2FV002627%2F1
MF, NR and~ST thank Google and the HEP Software Foundation for funding
via the 2020 and 2021 Google Summer of Code programmes. YY thanks the
Spreadbury Fund and the UCL Impact scheme for PhD studentship funding. JY
acknowledges an STFC doctoral studentship via the ScotDIST Centre for
Doctoral Training in Data-Intensive Science.

\clearpage
\bibliography{yoda2.bib}

\end{document}

%% file: preamble.tex
\hypersetup{
    colorlinks,
    linkcolor={red!50!black},
    citecolor={blue!50!black},
    urlcolor={blue!80!black}
}

\usepackage[bitstream-charter]{mathdesign}
\DeclareSymbolFont{usualmathcal}{OMS}{cmsy}{m}{n}
\DeclareSymbolFontAlphabet{\mathcal}{usualmathcal}

\fancypagestyle{SPstyle}{
\fancyhf{}
\lhead{\colorbox{scipostblue}{\bf \color{white} ~SciPost Physics Codebases }}
\rhead{{\bf \color{scipostdeepblue} ~Submission }}

\fancyfoot[C]{\textbf{\thepage}}
}

\usepackage{doi}

\usepackage{xspace}
\usepackage{graphicx}
\usepackage{afterpage}
\usepackage{lineno}
\usepackage{tikz}
\usetikzlibrary{arrows,shapes,backgrounds,fit,positioning,calc}
\definecolor{irn-bru}{RGB}{239,129,0}

\usepackage{subfig}
\usepackage{microtype}
\makeatletter
\g@addto@macro\bfseries{\boldmath}
\makeatother

\usepackage{xpatch}
\usepackage{xparse}

\usepackage{listings}
\lstMakeShortInline|

\usepackage{todonotes}
\makeatletter
\xpretocmd{\todo}{\@bsphack}{}{}
\xapptocmd{\todo}{\@esphack}{}{}
\makeatother

\usepackage{siunitx}
\DeclareSIUnit{\electronvolt}{\text{e\kern-0.15ex V}}
\DeclareSIUnit{\eV}{\electronvolt}
\DeclareSIUnit{\MeV}{\mega\eV}
\DeclareSIUnit{\GeV}{\giga\eV}
\DeclareSIUnit{\TeV}{\tera\kern-0.1ex\eV}
\DeclareSIUnit{\ifb}{\femto\barn\tothe{-1}}

\newcommand{\sw}[1]{\text{\textsf{#1}}\xspace}
\newcommand{\cxx}{\text{C\protect\raisebox{0.15ex}{\scalebox{0.8}{++}}}\xspace}

\newcommand{\python}{\text{Python}\xspace}
\newcommand{\cython}{\text{Cython}\xspace}
\newcommand{\numpy}{\sw{numpy}\xspace}
\newcommand{\mpl}{\sw{Matplotlib}\xspace}

\NewDocumentCommand{\rivet}{o}{\text{R\protect\scalebox{0.8}{IVET}\IfNoValueTF{#1}{}{\,#1}}\xspace}
\NewDocumentCommand{\contur}{o}{\text{C\protect\scalebox{0.8}{ONTUR}\IfNoValueTF{#1}{}{\,#1}}\xspace}
\NewDocumentCommand{\yoda}{o}{\text{Y\protect\scalebox{0.8}{ODA}\IfNoValueTF{#1}{}{\,#1}}\xspace}
\newcommand{\hepdata}{HEPD\protect\scalebox{0.8}{ATA}\xspace}
\newcommand{\ROOT}{\sw{ROOT}\xspace}
\newcommand{\boosth}{\sw{Boost.Histogram}\xspace}

\makeatletter
\let\@oldto\to
\renewcommand{\to}{\ensuremath{\@oldto}\xspace}
\renewcommand{\to}{\ensuremath{\@oldto}\xspace}

\makeatother

\renewcommand{\vec}[1]{\mathbf{#1}}
\renewcommand{\d}{\mathrm{d}}

\newcommand{\varspace}{\ensuremath{\Omega}\xspace}
\NewDocumentCommand{\varcoord}{o}{\ensuremath{\IfNoValueTF{#1}{\vec{\omega}}{\omega^{(#1)}}}\xspace}
\newcommand{\fillspace}{\varspace}
\let\fillcoord\varcoord
\newcommand{\binspace}{\ensuremath{\Theta}\xspace}
\NewDocumentCommand{\bincoord}{o}{\ensuremath{\IfNoValueTF{#1}{\vec{\theta}}{\theta^{(#1)}}}\xspace}
\newcommand{\unbinspace}{\ensuremath{\Upsilon}\xspace}
\NewDocumentCommand{\unbincoord}{o}{\ensuremath{\IfNoValueTF{#1}{y}{y}}\xspace} %< note difference

\newcommand{\defn}[1]{\text{\emph{#1}}\xspace}

%% file: sec-intro.tex
\section{Introduction}
\label{sec:intro}

In the current era of advanced statistical analysis methods and
toolkits -- from Bayesian inference to machine-learning -- the
simple concept of a histogram is often taken for granted. Yet it is
worth reflecting on how powerful a tool they are: a set of summary
statistics grouped into binned ranges of an independent variable or
variables, with a fixed data size and a mathematical definition
directly linked to differential and integral calculus. There is
surprising conceptual depth in these apparently simple objects.

The first of these features means histograms are an extremely
memory-efficient approach to approximating distributions. Unlike
``ntuple'' datasets, histogram objects are the same size regardless of
how many ``fill'' events are aggregated into them. As CPU capability
has grown much faster than RAM, this is an increasingly rather than
decreasingly important feature. The mathematical foundations
furthermore mean that the aggregation can be partitioned, allowing
parallel operation and/or continuous updating: particularly useful and
important features for processing of datasets much larger than can be
held in memory. Such applications abound in the modern era, from event
analysis at particle-physics facilities to analysis of datasets from
hundreds of millions of digital-service users.

Unfortunately, few computational libraries implementing histograms for
statistical analysis make full use of these parallels, which limits
functionality and can even lead to numerical approximations~\cite{roott2profile}.
%% and in so doing limit themselves to partial functionality,
%% inconsistencies, or broken metaphors.
The \yoda statistical analysis library is an outlier in this, having
been designed around a mixture of mathematical consistency and a
user-friendly programmatic interface. However, in the decade since the
release of \yoda~v1.0, limitations in that design also became
apparent, leading to a ground-up redesign to generalise the ideas of
the first version while respecting real-world requirements. This paper
first introduces the mathematical foundations of general statistical
histogramming, the resulting top-down requirements and motivations in
library design, then the specific approaches taken to achieve them in
(primarily) modern \cxx code and supporting tools.

%% file: sec-stats.tex
\section{Statistical preliminaries}
\label{sec:maths}
\label{sec:stats}

The ubiquity of histograms means it is typical to assume understanding
of what they are, which can lead to issues of mismatch between design
and usage. The most common such issue is a conflation between the
statistical content of a histogram bin and its rendering (e.g.~as bin
height) in a plotted representation. To avoid such ambiguities, we
first define the statistical objects of interest, before discussing
the software design constraints and the technical implementation.

\subsection{Unweighted mean and variance}

We start with the first and second-order \defn{statistical moments},
i.e.~the mean and variance, of an unbinned scalar quantity $x$, as
obtained from the probability density function (pdf) $f(x) \equiv
\d{P}/\d{x}$:
\begin{align}
  \label{eq:pdfmoments}
  \langle x \rangle &\equiv \int_{x \in X} x f(x) \, \d{x}\\[1em]
  \langle x^2 \rangle &\equiv \int_{x \in X} x^2 f(x) \, \d{x}\\[1em]
  \sigma^2(x) &\equiv \langle x^2 \rangle - \langle x \rangle^2 \, .
\end{align}
Of course these familiar quantities capture the average position and
the typical (squared) dispersion of the $f(x)$ distribution, with
perfect differential information about the density function. In
reality we are typically limited in our knowledge about such such
distributions as they are known not via an analytic form amenable to
integration, but via a finite-sized \defn{sample} from the
distribution (typically obtained either by experiments on natural
systems or by some computer code sampling the implicit distribution,
e.g.~a Monte Carlo event generator). From the finite sample we can
build estimators to these quantities, computed over a sampled set of
$N$ \defn{fill values} $\{ x_n \}$ for $1 \le n \le N$, and unbiased
in the sense that they converge to the ideal values in the limit of
infinite sample size, $N \to \infty$:
\begin{align}
  \label{eq:unweighted_mean}
  \langle \hat{x} \rangle_\mathrm{U}
  &\equiv \frac{\sum_{n=1}^N x_n}{N}\\[1em]
  \label{eq:unweighted_var}
  \hat\sigma_\mathrm{U}^2(\hat{x})
  &\equiv \frac{\sum_{n=1}^N (x_n - \langle x \rangle)^2}{N-1}\nonumber\\
  &= \langle x^2 \rangle_\mathrm{U} - \langle x \rangle_\mathrm{U}^2\nonumber\\
  &= \frac{\sum_{n=1}^N x_n^2}{N-1} - \frac{\left( \sum_{n=1}^N x_n \right)^2}{(N-1)^2} \, .
\end{align}
Here the $N-1$ term in the variance is the Bessel correction to
achieve an unbiased estimator, one degree of freedom having already
been used to obtain the estimated mean.

\subsection{Estimated counts and efficiencies}

A closely related quantity is the Poisson estimator of the uncertainty
on the \defn{count} of a random variable, i.e.~the number of observed
fills in a data-sampling run of fixed length. The variance of a
Poisson random variable
%% (such as the fill-rate of a counter, over a defined period)
is equal to its mean, and as our best estimate of the Poisson mean is
the final count, the best estimate of the Poisson variance is also the
count: $\sigma^2(N)_\mathrm{P} = N$. As a result, the relative
uncertainty on the count decreases with the classic Monte Carlo
scaling, $\sigma(N)/N = 1/\sqrt{N}$.

Poisson rate is distinct from the \defn{efficiency} of a fill, which
is the fraction $\epsilon$ of a known number of total events $N$ to
pass some requirements. If $N_\mathrm{sel}$ of the total $N$ are
selected, the sample efficiency is defined as $\hat\epsilon \equiv
N_\mathrm{sel}/N$. An analytic estimator of the uncertainty on the
efficiency is obtainable via Binomial statistics:\footnote{As opposed to
Poisson, since the number is now fixed and we are choosing between
selected or not, rather than an unknown total number with known rate
of random occurrence.}
\begin{equation}
  \hat\sigma^2(\hat\epsilon)_\mathrm{B}
  = \frac{\hat\epsilon (1 - \hat\epsilon)}{N} \, ,
\end{equation}
which has the pleasing feature that the one-sigma error band
$[\hat\epsilon - \sigma(\hat\epsilon)_\mathrm{B} \ldots \hat\epsilon +
  \sigma(\hat\epsilon)_\mathrm{B}]$ is always contained in the range
$[0,1]$: this is semi-accidental, but conveniently logical, and
sufficient for most practical purposes. (In the rare cases where it is
not, a non-analytic MC-toy computation of the efficiency probability
distribution is usually required.)

\subsection{Weighted moments, mean and variance}

Returning to our weighted moments, we can extend the concept of our
finite set of sampled values to include the concept of \defn{fill
  weights}, $w_n$. These represent rescalings of the probability of
the fill with respect to the nominal $w_n = 1 \, \forall n$ implicit in
the unweighted case: if $w_n > 1$ it indicates that this single fill
is statistically representative of greater than one fill's worth of
probability density, and \emph{vice versa} if $w_n < 1$ it indicates
that this fill-value $x_n$ (or at least the route by which it was
obtained) has been overrepresented and should be considered as worth
less than one fill.

Weights can occur either for natural measurements or for simulated
ones, reflecting either or both of a biased selection process or a
variation to assumptions made in the measurement process.  For
example, on the first point, a measurement may use ``prescaling'' to
randomly discard events/fills selected by different routes in order to
limit data rates and balance contributions from the different sources,
or an MC simulator may intentionally deviate its samplings from the
naive probability distributions in order to again achieve a more
equitable distribution of fill counts across the space of values~\cite{2021pcp}. On
the second point, the values being filled may have been obtained by
inference from a more fundamental set of ``raw'' observations: if the
nominal assumptions underlying that inference are varied, the relative
probability of each fill-value will change, giving a weight $w \propto
P_\mathrm{var}/P_\mathrm{nom}$.

In short, fill-weights are a mechanism for offsetting biases in the
fill-generation process in order to obtain an unbiased distribution or
its moments. Hence, we must generalise our moment estimators to
include the effects of fill weights (where from now on we drop the
explicit summation ranges over sample indices):
% See: https://stats.stackexchange.com/questions/47325/bias-correction-in-weighted-variance
%
\begin{align}
  \label{eq:weighted_mean}
  \langle \hat{x} \rangle
  &= \frac{ \sum_n w_n x_n }{ \sum_n w_n } \\[1em]
  \label{eq:weighted_var}
  \hat\sigma^2(\hat{x})
  &= \mathcal{B} \cdot \frac{ \sum_n w_n \left(x_n - \sum_m w_m x_m \right)^2 }{\sum_n w_n}\nonumber\\
  &= \mathcal{B} \cdot \frac{ \left( \sum_n w_n x_n^2 \right) \cdot \left( \sum_n w_n \right) - \left( \sum_n w_n x_n \right)^2 }{ \left( \sum_n w_n \right)^2 }\nonumber\\
  &= \frac{ \left( \sum_n w_n x_n^2 \right) \cdot \left( \sum_n w_n \right) - \left( \sum_n w_n x_n \right)^2 }{ \left( \sum_n w_n \right)^2 - \sum_n w_n^2 } \, .
\end{align}
As for the unweighted cases, these expressions incorporate the effect
of Bessel's correction, subtracting ``one fill's worth'' of
information in the computation of the outer expectation values. This
correction is captured in the multiplicative factor $\mathcal{B}$,
which for unweighted statistics is $\mathcal{B}_\mathrm{U} = N/(N-1)$
to effectively replace the na\"ive $1/N$ averaging with $1/(N-1)$. For
weighted statistics what ``one fill'' corresponds to is encoded in the
\defn{effective fill-count},
\begin{equation}
  N_\mathrm{eff} = \frac{ \left( \sum_n w_n \right)^2 }{ \sum_n w_n^2 } \, ,
\end{equation}
which is used by direct replacement to give the \defn{weighted Bessel factor},
\begin{align}
  \label{eq:weighted_bessel}
  \mathcal{B}
  &\equiv \frac{N_\mathrm{eff}}{(N_\mathrm{eff}-1)}\nonumber\\
  &= \frac{\left( \sum_n w_n \right)^2}{\left( \sum_n w_n \right)^2 - \sum_n w_n^2} \, ,
\end{align}
and hence the final form in eq.~\eqref{eq:weighted_var}. Note that the
effective replacement on the denominator is now $\left(\sum_n w_n
\right)^2 \to \left( \sum_n w_n \right)^2 - \sum_n w_n^2$.  It is
simple to verify that the effective count is invariant under uniform
global rescalings of all weights, $w_n \to a w_n$, and hence if all
fills have the same weight, however large or small that weight is, the
effective fill-count of the sample is equal to its na\"ive number of
entries, $N$. In weighted statistics, $N_\mathrm{eff}$ indicates the
degree of statistical stability of the sample, which can only be
smaller than in the unweighted (or equivalently, equally weighted)
case. In extreme circumstances such as balanced fills with positive
and negative weight-sign, the effective number of fills can be zero,
as a useful indication that the variance estimate will be unreliable.

%% \begin{equation}
%%   \sigma^2(x) = \frac{ \left( \sum_n w_n x_n \right)^2 }{ \sum_n w_n }
%% \end{equation}

\subsection{Histograms}

Having established a coherent statistical picture for unbinned
quantities in one dimension, we now both generalise the measured
variable $x$ to a vector \defn{variable-space} \varspace composed of
vectors \varcoord and with a differential volume element
$\d{\varspace}$, and partition that space into a disjoint (sub)set of
\defn{bins}, $\{ \varspace_b \} \subset \varspace$.

The generalisation of the variable dimensionality has little profound
effect beyond extending the possible set of weighted moments to
$\langle \varcoord[i] \rangle$ and $\langle \varcoord[i] \varcoord[j]
\rangle$ for all dimension indices $i,j$ in \varspace.  In general the
cross-terms of this generalised second moment encode variable
correlations, which can be transformed into a sample-covariance
matrix via generalisation of the moment-construction for self-variance,
including the weighted Bessel factor of eq.~\eqref{eq:weighted_bessel}:
\begin{align}
  \widehat\Sigma_{ij}
  &= \langle \varcoord[i] \varcoord[j] \rangle - \langle \varcoord[i] \rangle \langle \varcoord[j] \rangle\nonumber\\
  &= \mathcal{B} \cdot \frac{\left( \sum_n w_n \varcoord[i]_n \varcoord[j]_n \right) \cdot \Big( \sum_n w_n \Big) - \left( \sum_n w_n \varcoord[i]_n \right) \cdot \left( \sum_m w_m \varcoord[j]_m \right)}{\Big( \sum_n w_n \Big)^2}\nonumber\\
  &= \frac{\left( \sum_n w_n \varcoord[i]_n \varcoord[j]_n \right) \cdot \Big( \sum_n w_n \Big) - \left( \sum_n w_n \varcoord[i]_n \right) \cdot \left( \sum_m w_m \varcoord[j]_m \right)}{\Big( \sum_n w_n \Big)^2 - \sum_n w_n^2} \, .
\end{align}

The bin partitioning introduces a new concept: where the unbinned
moments converged to summary statistics of the entire probability
density function $f(\varspace) \equiv \d{P}/\d{\varspace}$, the
moments in each bin $b$ converge to the summary properties of that
bin's variable-space partition,~i.e.
\begin{align}
  \langle \varcoord[i] \rangle_b &\equiv \int_{\varcoord \in \varspace_b} \varcoord[i] f(\varcoord) \, \d{\varspace}\\[1em]
  \langle \varcoord[i] \varcoord[j] \rangle_b &\equiv \int_{\varcoord \in \varspace_b} \varcoord[i] \varcoord[j] f(\varcoord) \, \d{\varspace} \, .
\end{align}
For consistency, all moments need to converge to their unbinned values
when the partition is expanded to include the whole space, and to
converge to the differential properties of the probability density
function itself when the partitions are made infinitesimally small,
$\varspace_b \to \d{\varspace}(\varcoord)$. In addition, integrating
out dimensions of the variable space (i.e.~combining bins along axes)
must for consistency converge to the same result as having originally
constructed a lower-dimensional or less finely binned partition of the
space: this is guaranteed by the linearity of the sums over the sample
indices in the statistical-moment definitions.

This perspective illustrates the fundamental connection between
differential histogramming and differential calculus: a statistical
histogram is not just a collection of fill counts, but a discrete
approximation to a continuous probability or population
distribution. The ideal differential distribution (or density) is
$\d{P}/\d{\varspace}$ in the case of a probability density, and
$\d{N}/\d{\varspace}$ in the case of a population density; histograms
approximate this by converting the differentials to finite deltas,
$\Delta{P}/\Delta{\varspace}$ and $\Delta{N}/\Delta{\varspace}$,
though the differential analogy is often preserved by use of the
infinitesimal $\d$ symbol in plot labelling.

We note in particular that the \defn{bin measure} $\d{\varspace}$ or
$\Delta\varspace$ representing the volume element of the bin is a
crucial consistency element in constructing a histogram's bin values:
to preserve the density estimate, the width (or generally volume)
$\mathrm{d}\varspace$ of the containing bin must be divided out so the
$\Delta \to \mathrm{d}$ limit converges. This is particularly
important as in general it is not desirable for finite bins to have
the same width: for the statistical relative uncertainty on bin
populations to be equally distributed across the histogram, bins
\emph{should} be made larger in regions of low density, and narrower
(until the variable-resolution limit) where there is high sample
density. With non-uniform bin sizes, failing to divide by the bin
measure distorts the distribution away from its physical
shape.

Should one wish to compute and display the actual bin population, one
would need to either -- ideally -- use a discrete binning expressed in
terms of finite probabilities rather than densities, or as a
workaround multiply each density bin by its fill-volume. In the
absence of a more official name for this object, and reflecting its
typical use, we refer to this not as a histogram, but as a \defn{bar
  chart}.

%% In addition, not all bins need be of the same width: for statistical
%% stability of estimated distributions one typically wants to achieve
%% a similar population in each bin, meaning their widths (or volumes)
%% should be distributed similarly to the probability density
%% $f(\vec{x}) \equiv \mathrm{d}P(x)/\mathrm{d}\vec{x}$ of the sampling
%% function. For a distribution to be asymptotically invariant under
%% choices of relative bin volume, the value needs to be scaled by the
%% reciprocal of the volume, as in the differential symbol.

\subsection{Profiles}

Our final statistical preliminary is to slightly generalise this
concept of a histogram. In the previous section we defined the binning
partition across the entire variable-space \varspace. But a useful
class of histogram mixes binned and unbinned variable subspaces,
allowing characterisation of the unbinned dimensions \unbinspace via
their moments as projected into each partition of the
\defn{bin-space} \binspace.

These partially binned objects are known as \defn{profiles},
effectively histograms in the bin-space with augmented moment content
in each bin. Conceptually, they are very useful objects for scientific
work as they allow statistical aggregation of finite samples into
``independent variable'' bins $\bincoord \in \binspace_b$, while
characterising the mean dependence of the unbinned dependent variables
$\vec{\unbincoord}$ on $\bincoord$.  Again, the limiting behaviour
must be that for infinitesimally small bins and infinite sample
statistics, the fully differential relationship
$\vec{\unbincoord}(\bincoord)$ is obtained. Aggregation of bins and
reduction of the unbinned space to lower-dimensions must again, as
with aggregation of the binned subspace, give the same result as
having originally constructed the lower-dimensional profile -- again
guaranteed by the linearity of the statistical moments.

In general a profile's unbinned space, \unbinspace, can be
multidimensional, but this introduces an ambiguity as to the resulting
canonical bin value -- each profile bin effectively contains the
histogram-type integrals of fill weights as well as the set of moments
corresponding to a multivariate Gaussian\footnote{Or more generally an
elliptical distribution.} distribution between the unbinned
variables. For definiteness, we restrict ourselves to a
single-dimensional unbinned space \unbincoord, whose relevant moments
are $\langle \unbincoord \rangle$ and $\langle \unbincoord^2 \rangle$.
Conventionally the profile canonical bin value is the mean $\langle
\unbincoord(\binspace) \rangle$ as a function of the binned
coordinates, and rather than the standard deviation of the unbinned
distribution, the \defn{standard error}
$\hat\sigma_{\bar\unbincoord}(\bincoord) = \hat\sigma_b / \sqrt{N_b}$
is used as the nominal uncertainty, where $N_b$ is the effective
sample count in bin $b \supset \bincoord$.

%% ``In (particle-)physics measurements, the ultimate intention is often
%% to achieve the best estimate of a ``true'' physical distribution or
%% dependence, independent of the particular apparatus used for the
%% measurement. One might, for example, want to estimate a probability
%% density $f(x) = \mathrm{d}P/\mathrm{d}x$ distributed in some
%% observable $x$, or characterise the dependency of one quantity on
%% another, e.g.~the dependence of a mean current on the controlling
%% voltage, $\langle I \rangle(V)$.''

With this, we have finished setting the scene: while not complex in
grand terms, the importance of consistency constraints and the
potential for high-dimensional and combinatoric complexity is
clear. In the next section we consider the implications for a
computational implementation that preserves these principles, and
establish core design goals for an implementation.

%% file: sec-design.tex
\section{Design principles}
\label{sec:design}

\yoda is a package for creation and analysis of statistical data,
particularly various flavours of histogram, written primarily in \cxx
and programmatically usable from \cxx and \python. The development and
developers of \yoda emerged from the sub-field of Monte Carlo event
generator analysis and tuning~\cite{Buckley:2011ms} in high-energy
physics (HEP), which places certain requirements and emphases on its
functionalities, but the library is deliberately agnostic of any
particular application.

A brief list of these requirements is useful to frame our design
choices, though we will see that some of these ideas naturally led to
generalisations beyond what we present here:
\begin{description}
  \item[Differential consistency:] a histogram is fundamentally not just
  a list of (weighted) fill-counts, but a binned best-estimate of a
  continuous distribution. Optimal estimation requires non-uniform
  binnings in proportion to expected probability density, and so it is
  crucial that these different fill-space volumes be divided out for
  each bin when estimating the binned density, i.e.~taking the
  $f(\vec{x}) \equiv \mathrm{d}P(x)/\mathrm{d}\vec{x}$ notation
  literally.

\item[Continuous aggregation:] for studying extremely large datasets
  as is often the case in HEP, it is not feasible to perform
  histogramming in one pass over a full set of fill values stored in
  memory. Over thousands of fill observables and potentially billions
  of events, this mode (as exemplified by e.g.~\numpy and \sw{Matlab})
  is not computationally viable. Instead, many HEP and similar
  large-data applications require a mode
  %% the established mode (familiar to most through the ROOT library)
  in which histograms are ``live'' summary objects to which new
  entries/fills can be continuously added.
  %% containing update-able variables to which new
  %% fills (or other aggregated summary objects) can be aggregated.
  
\item[Weighted statistical moments:] the key summary statistics needed
  from histogram bins are the bin value (as defined by the type of
  histogram being used) and its statistical uncertainty, but also the
  mean and (co)variance of the fill-distribution within the bin's
  coordinate space. The fills are in general weighted according either
  to how they were sampled or manipulated, and so histograms need to
  store the weighted statistical moments required to compute the key
  summary statistics of their bins.

  This picture also provides a consistent way to view the extended
  ``profile'' histogram type: in addition to storing the statistical
  moments of the fill weights and the fill dimensions, a profile bin
  also stores the moments of further unbinned dependent values,
  $\vec{y}$.
  %% not (just) the raw probability distribution,
  %% but of a further dependent variable $I$ as a function of the binned fill coordinates

\item[Integral consistency:]
  Statistical moments, as defined from the pdf function in
  eq.~\eqref{eq:pdfmoments}, are intrinsically integral quantities,
  computed via marginalisation across the variable-space $\varspace$,
  or subspaces of it. This integral property maps into the sampled
  moments and summary estimators. As such it should be possible to
  reconstruct their full integral values (or the best available
  estimate from finite fill statistics across the whole space) by
  composing together their estimates in subspaces, e.g.~in different
  bins. A consistent computational statistics library using binned
  quantities should be able to project high-dimensional binnings into
  lower-dimensional ones (for example, constructing a binned profile
  along one axis of a higher-dimensional histogram) without biasing
  integral quantities, e.g.~by replacing them with discretised
  bin-centre estimates in the high-dimensional binning.
    
\item[Separation of style from substance:] a histogram is first and
  foremost a data object representing the statistical properties of
  the fills recorded into its bins, rather than any particular
  rendered representation of that data. One should be able to ensure
  the invariance of statistical data while varying the plotting style.

\item[Separation of binning from bin-content:] this latter point can
  be more generally viewed as motivating a separation between
  \defn{live} and \defn{inert} classes of data-object: the former are
  the statistical objects tracking the evolution of summary moments
  (and permitting further data-taking) %but without specifying a
  % visualisation to be derived from that binned data,
  while the latter
  are a specific representation of chosen data facets into a set of
  ``values and uncertainties'' for unambiguous plotting or
  summarising.  But once this thought-process is underway, a further
  design clarification arises: the mechanism for aggregating
  multi-dimensional coordinate ranges into discrete bins is useful
  independently of the statistical moments they contain, and a suitable
  generalisation makes it possible to unambiguously implement live and
  inert binned data types using the same abstract binning framework.

\item[User friendliness:] while any sort of programming would be
  regarded as user-unfriendly in many circumstances, programmatic
  data-analysis is the norm in quantitative science, and the
  programming interfaces of libraries cover a wide spectrum often
  between powerful-but-intimidating and easy-but-inflexible. Our
  motivation is to provide a ``clean'' programmatic interface
  expressed in terms of statistical and data-analytic concepts and
  hence well-matched to the goals and skill-sets of data scientists,
  rather than emphasising language technicalities or requiring
  systems-programmer levels of programming familiarity.
\end{description}

%% \TODO{Error propagation, variable bin widths, live and inert data
%%   types, projections and integrals, encapsulation and
%%   consistency/contract enforcement, global and local indexing,
%%   overflows}

The continuous-aggregation design principle excludes many existing
data-analysis libraries from contention for HEP and other large-data
tasks. The most prominent packages for continuous-aggregation
histogramming are the HEP \ROOT~\cite{Brun:1997pa} framework and
arguably the \boosth~\cite{boosth} library.

The former is heavily used in HEP for histogramming in low
dimensionality with limited axis and bin-content types, but conflates
live and inert types (allowing for inconsistency with fill history),
and data and presentation uses.\footnote{These should be read not as a
criticism, but as a pragmatic design choice on which \yoda takes a
different path.} \ROOT is also a very large monolithic framework with
many dependencies, including many features beyond statistical analysis
and expecting to be the controlling element in applications rather
than called as a utility library: these can be obstacles to use in
many applications.

By contrast, \boosth focuses on an abstract implementation of binning
in a style reminiscent of the \cxx standard library, and making
explicit use of advanced language features and abstractions distanced
from statistical terminology. It is hence a powerful tool, but the
interface imposes a barrier for less technically adept \cxx users. In
the design of \yoda[2] we attempt to negotiate a third path in which
advanced language features (similar to the \boosth approach) are
\emph{internally} used to enable high levels of abstraction and to
enforce statistical consistency and type-safety, but also providing
a \cxx (as well as \python) user-interface in which this complexity is
hidden and the application alignment emphasised. \yoda is also
intentionally limited to statistical analysis only, requires no
library dependencies for core \cxx operation, and operates purely as a
class library rather than a stateful controlling framework -- choices
explicitly made to assist embedding into applications.

%% file: sec-yoda1.tex
\section{Experience from \yoda version~1}
\label{sec:yoda1}

The top-level design goals stated above were partially established at
the time of the \yoda version~1.0 release in 2013, e.g.~with fully
correct (to second order) statistical-moment combination a core
principle of the design. But deployment experience in the meantime,
and in particular extension to multidimensional histograms binnings,
revealed structural issues which motivated the rewrite described in
the following section:
\begin{description}
\item[Wrapper-type issues:] A \emph{good} design decision was to
  isolate statistical moment calculations into a set of |Dbn|
  (distribution) classes for each dimensionality. For example, a
  weighted ``counter'' type was just a zero-dimensional distribution
  which tracked sums of weights and squared weights, each histogram
  bin was implemented around |Dbn1D|, one-dimensional profile
  histograms contained a |Dbn2D| in each bin, and so-on. But each of
  these bin-wrapper types required repeating and mapping the majority
  of |Dbn| interface methods, which with the combination of binning
  and |Dbn| dimensionalities created a significant maintenance
  overhead. The need for actual wrapper types to be stored also
  bloated the analysis-object memory requirements and fragmented the
  memory layout relative to a contiguous array of the fundamental
  |Dbn|s.
  
\item[General irregular binnings:] \yoda[1] had intrinsic, first-class
  support for non-uniform binnings, as opposed to an awkward add-on,
  as in many HEP applications the probability density over an
  observable's range of interest varies by orders of magnitude: with
  non-uniform binning as the only reasonable strategy to balance
  resolution with statistical precision, it needed to be easily
  invoked. But the general-dimensionality extension of non-uniform
  binnings is irregular tilings of the binning space, and here the
  ambition proved a step too far. Even in two dimensions, computing
  the (non-)overlap or commonality of two binnings was a significant
  computational challenge, and more complex in higher
  dimensionalities, for little or no practical gain.

\item[Overflows:] Noting the unphysicality of bin-property requests to
  underflows and overflows in the \ROOT histogram classes,
  \yoda[1] implemented overflow bins as bare |Dbn| types, avoiding
  interpretation as standard bins with widths/areas, etc. In
  one-dimensional data types this was no issue, as there were only two
  overflows to be dealt with, i.e.~ the underflow and overflow below
  and above the binned range respectively. But in 2D (and higher
  dimensionalities), distinct overflow distributions were needed above
  and below every row and column of the in-range binning: an
  exponentially more complex problem, and even an uncomputable one
  given the promise of general irregular binnings and gaps. \yoda[1]
  never provided fully functional overflows in multiple dimensions,
  blocking for example the dimensional reduction of a 2D histogram
  into a projected 1D histogram or profile.
  
\item[Bin gaps:] Motivated by gaps in binning\footnote{Not
just empty bins, but unreported regions in the fill range.} in legacy
  data records, \yoda[1]'s binnings were implemented as a list of
  explicit bin objects which knew their own edge locations,
  information then duplicated by the axis objects responsible for
  locating the bins given a set of fill coordinates. A hidden ``total
  distribution'' object was required to track fills that landed in the
  gaps but should still contribute to overall normalization. This
  structure also lended itself to ability to sequentially add and
  remove single bins, which then required a locking mechanism for
  consistency to ensure that binnings could not be altered once
  filling had begun. Again, this was significant unnecessary
  complexity and maintenance difficulty.
  
\item[Mismatched live/inert types:] \yoda[1] did clearly separate live
  histogram types from inert types, largely informed by necessity due
  to the limited information available in reference-data resources
  such as the \hepdata publication-data database. But all inert types
  were implemented as ``scatter'' objects: a set of $n$-dimensional
  points and error bars, without any concept of binning. This made
  comparisons between inert scatter types and live types awkward, with
  the binning needing to be heuristically reconstructed from the
  absolute positions of the ends of the error bars, for example if
  using a reference-data object to set the appropriate set of bin
  edges on a new live object. This process worked in most
  circumstances, but was not efficient, and was based on guesswork and
  convention rather than a type-safe approach. In addition, as an
  inert format is naturally the type to be used for data preservation
  and interpretation, more structured statistical data such as
  uncertainty breakdowns (for various classes of statistical or
  systematic error) was retrofitted incoherently on to the scatter
  types. This involved more arbitrary or conventional heuristics,
  rather than a fully coherent design, and opportunities for
  inconsistency in conversion between pointwise and whole-object views
  of the encoded correlations.
\end{description}

%% Despite our best efforts, YODA v1 also has several shortcomings, for example:

%% - limitation of data-object dimensionality, and to continous-valued axes;

%% - inability to store arbitrary data-types in binnings;

%% - correct but limited treatment of overflow bins, particularly in >1 dimension;

%% - no unified scheme for local and global bin indexing in multiple dimensions;

%% - internal code duplication to support C++ and Python APIs for several different
%%   dimensionalities and binned-content types;

%% - mismatching of the "inert" scatterplot data type from e.g. HepData to the
%%   binned "live" objects from MC runs;

%% - limited and inconvenient implementation of multiple error-sources and
%%   error correlations on the scatter types;

%% - the choice of a (possibly compressed) plain-text data format does not
%%   scale well to multiweighted simulations or HPC operation.

%% file: sec-impl.tex
\section{Implementation}
\label{sec:impl}

All these issues, and myriad smaller ones, led to a wholesale rethink
of the \yoda implementation in order to better meet the now clear
requirements of a general and fully consistent histogramming
library. At the same time, it remained paramount that the resulting
API be usable by a typical physics programmer, via compact and
expressive data-handling code either in straightforward \cxx or via the
\python wrapper package.

As the mathematical principles of consistent multidimensional
statistics, and the resulting design goals of the redesign, are
expressible declaratively at compile time, there is a natural synergy
between this application and the template-metaprogramming methods of
modern \cxx. As well as guaranteeing conceptual consistency and type
correctness, and avoiding the code-duplication maintenance woes of the
original \yoda release series, a template-oriented approach is
consistent with engineering for high-performance applications, rather
than deferring dimensionality calculations to runtime. Accordingly,
the \yoda[2] architecture is heavily based on modern \cxx17 template
methods~\cite{10.5555/2685398}, and will evolve in-line with the
language's focus on making such methods more powerful and
accessible~\cite{ISO:2017:IIIa}.

\subsection{Bin partitioning}

\yoda[2] builds its partitionings of the fill space by
%an ``outer product''
composition of multiple one-dimensional binnings. These 1D binnings
are implemented via a unified |Axis| class (a refinement of \yoda[1]'s
|Axis1D|) that is templated on the edge type. The default
implementation assumes a discrete binning with one bin for each edge
label and an additional catch all-else ``otherflow'' bin at index |0|,
while the more traditional continuous-value axes are provided as a
template specialisation triggered when the template type satisfies the
|std::is_floating_point| trait.

The floating-point specialisation implements $N-1$ bins for $N$ edges,
as each bin is enclosed by a lower and upper bin edge with a shared
edge between two neighbouring bins. It also adds an underflow bin
between |-inf| and the lowest finite edge as well as an overflow bin
between the highest finite edge and |+inf|. This infinite-range
binning ensures consistent slicing behaviour in higher dimensions
across multiple axes and their respective sets of under- and overflow
bins. \yoda actively uses the IEEE\,754 floating-point
standard's~\cite{1985--ieee754} |inf| and |nan| values, and their
standard combination rules with normal floating-point values to
express the range of computable binning and bin-value quantities
rather than necessarily treating these overflow values as errors.

Helper methods are provided for 1D binning-specification, including
\sw{Matlab}/\numpy-influenced |linspace| and |logspace| functions,
and a general |pdfspace| for distribution of $N$ bins proportional to
an arbitrary probability density.

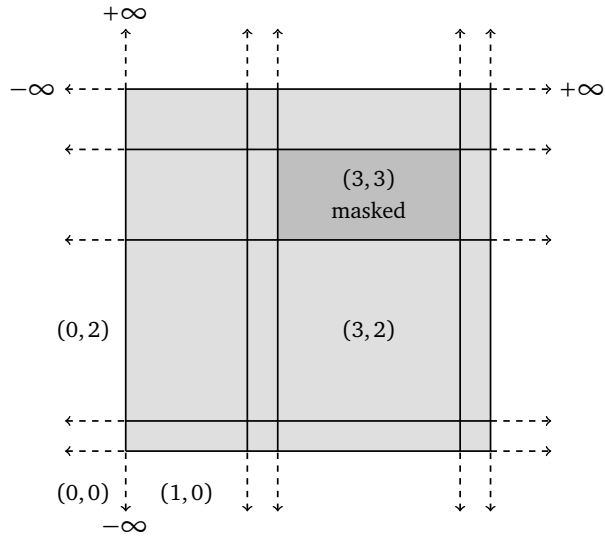
\begin{figure}%[htb!]
 \input{binning-diagram}
 \caption{Illustration of the \yoda[2] generalised infinity-binning
   scheme for continuous axes. The grey-shaded region represents
   in-range bins, with the white regions extending to positive and
   negative infinity being the overflow bins. Local-index tuples are
   annotated on several bins, illustrating the zero-indexing of
   axis-local indices when including underflow and overflow
   bins. %% Masking of a bin, which remains in the binning but is simply
   %% disabled from plotting and statistics computations, 
   Bin masking is shown at index $(3,3)$.
   %% This scheme is demonstrated
   %% in two dimensions for clarity, but generalises to arbitrary
   %% dimensionalities with a mixture of discrete- and continuous-typed
   %% binning axes.
 }
 \label{fig:binning-diagram}
\end{figure}

The global fill-space $\fillspace$ is partitioned into bins via the
outer product of independent 1D axes, giving a rectilinear (but
generally non-uniform) binning grid as illustrated in
Figure~\ref{fig:binning-diagram}. A single bin on one independent axis
corresponds to a family of bins in the remaining dimensions of the
fill-space: These mappings can be efficiently computed, permitting
general \defn{slicing} and \defn{marginalising} across global
fill-space.

This functionality is implemented in a dedicated |Binning|
class, which can also translate between the tuple of local bin indices
(one for each axis) and a global bin index (in $\fillspace$-space) via
|globalToLocalIndices| and |localToGlobalIndex| methods, using a
standard nested-stride mapping scheme.

The treatment of underflows and overflows as fully fledged bins means
that the local indices start at |1| for the in-range bins. Gaps in
fill-space are supported via \emph{bin masking} rather than bin
erasure, ensuring that the underlying binning always counts all fills
and that total fill statistics can be computed by integration over
bins, rather than losing information in bin-gaps (or needing to
maintain a separate ``total'' counter as in \yoda[1]).

This simple factorisation of per-axis binning structures avoids the
issues with high-dimensional bin-overlap computation and necessity of
fill-locking experienced with \yoda[1]'s na\"ively ambitious
arbitrary-binning model. Should a user wish to use irregular bin
tilings across the global fill-space, these are not directly
supported, but can be achieved via post-processing given a suitably
granular initial binning.

The factorised-binning model is also efficient in regular use, as the
global bin-index is computed in fixed time from the set of local
bin-indices along each axis; these in turn are computed in an
optimised fashion using a |BinEstimator| object created along with the
binning. This pre-assesses whether a linear search strategy, or a
binary-search in either linear or log-space will be the most efficient
for bin discovery given a fill coordinate, such that the asymptotic
complexity of $D$-dimensional bin-lookup over $N_\mathrm{bin}$ bins
per axis is $\sim D \ln N_\mathrm{bin}$. For use-cases where the same
coordinates will be used repeatedly on equivalent histograms, the bin
index can be cached to avoid duplicated bin-search computations.  The
global index can then be computed and used in $\mathcal{O}(1)$ time
for access to contiguous storage corresponding to the bin contents.

\subsection{Bin content}

The \emph{content} of the partitioned bins is generalised in \yoda[2]
to allow storage of arbitrary types within binnings, paving the way
for discrimination between simple storage of inert quantities (e.g.~a
float-valued binned efficiency-map) and more tightly integrated live
types as discussed in Sections~\ref{sec:design} and~\ref{sec:yoda1}.

To maintain the impression of bins as distinct objects looked-up
within the binning, and with awareness of their location as well as
content, a templated |Bin| class is provided to wrap around the bin
content (by inheriting from and augmenting it with bin-location
information such as indices, edges and geometric midpoints) and to
provide a link to the parent |Binning| object.

Any given bin covers a volume element |dVol()| in $\fillspace$ fill
space, given by the product of all continuous-axis bin widths, which
can become infinite if a bin falls into the under- or overflow range
of a continuous axis. Semantic |dLen()| and |dArea()| aliases are
provided in 1D and 2D, respectively. In mixed continuous--discrete fill
space, the discrete bins default to a unit width to allow for the
fill-space element to be finite.  The concept of ``height'' that was
present in \yoda[1], motivated by typical graphical histogram
representations, has been removed from v2 as there is neither a
guarantee that the stored content is arithmetic, nor that the
traditional plot representation is relevant.

Access to axis-specific quantities is via templated accessor methods,
e.g.~a bin |b|'s minimum-value edge on the $d$'th axis is accessed like
|b.min<$d-1$>()|, noting the zero-indexing of the axis indices. The
curiously recurring template pattern (CRTP) is used to mix in
axis-specific method names for the first three dimensions (|xMin()|,
|yMax()|, etc.)  and to reduce the amount of code duplication.

\subsubsection{Live content}
The |Dbn| distribution class from \yoda[1] has been generalised to
arbitrary dimensions by templating it on the dimensionality. It tracks
the number of events, the sums of weights and squared weights, and the
weighted first- and second-order moments of axis position (including
mixed moments e.g.~$\sum_n w_n x_n y_n$) which allow computation of
the means and variances along each independent axis as well as the
general covariance matrix. The vector mean position of weighted fills
within a bin is referred to as the \defn{bin focus} and is exposed
through the bin interface, meaning that there is no need to
approximate a bin's effective statistical location as being at its
geometric midpoint.

The |Dbn| class has a |fill| method that accepts the next weight, a
coordinate for each dimension, and an optional \emph{fill
fraction}. The concept of a fractional fill is motivated by numerical
instabilities in quantum chromodynamics, where the ``perfect
resolution'' of a sharp bin-edge can introduce non-cancellation of
infinities (or in computational reality, non-cancellation of large,
oppositely-signed weights in correlated fills); by spreading a fill
over neighbouring bins, numerical stability can be restored. While
motivated by quantum-mechanical issues, the additional concept of a
fill-fraction in addition to the more established fill-weights may be
of use in other domains.
%% relevant in some next-to-leading-order QCD generators, where e.g.~in a
%% dipole subtraction scheme~\cite{Catani:1996jh}, real emission events
%% are produced as a set of events that include correlated but negatively
%% weighted counter-events. These should in general be treated as single
%% events, with the event weight given by the sum of weights of the real
%% and counter events in the event group, giving rise to fractional fills
%% within an event group.

\subsubsection{Inert content}
Representation of binned inert content is a new feature in \yoda[2],
reflecting the conceptual mismatch in \yoda[1] between binned live
types and point-based representation of inert reference-data.

A new |Estimate| class has been introduced, consisting of a single
central value as well as an optional dictionary of labelled error
pairs, representing identified sources of nominally 1-sigma
uncertainty (from which a systematic covariance can be constructed
between bins). The signed error pairs are understood to be the results
of respectively downward or upward shifts in a correspondingly named
nuisance parameter; the empty string is interpreted as a user-supplied
total uncertainty pair. Error labels starting with |stat| or |uncor|
are treated as uncorrelated in arithmetic operations, while all other
error labels result in a fully correlated treatment of the
corresponding error pairs. If a different behaviour is desired, an
alternative |regex| parameter can be provided to processing functions
such as |divide|, to trigger the uncorrelated treatment.  Various
convenience methods are provided to translate the default down/up
representation into generally asymmetric negative/positive error bars
induced on the estimate value.  The total uncertainty is by default
given by the sum-in-quadrature of all negative and positive
components, unless a user-provided total uncertainty pair is present.

\subsection{Combined partitioning and content}

\begin{figure}%[htb!]
 \input{storage-diagram}
 \caption{Illustration of the internal \texttt{BinnedStorage}
   structure in \yoda[2]. Each bin is a light wrapping of the raw
   content type, with bin-location information made accessible via
   mixed-in methods making use of the linked \texttt{Binning}
   object. Methods implemented on \texttt{BinnedStorage} may rely on
   either the binning data, the content data, or both.}
 \label{fig:storage-diagram}
\end{figure}
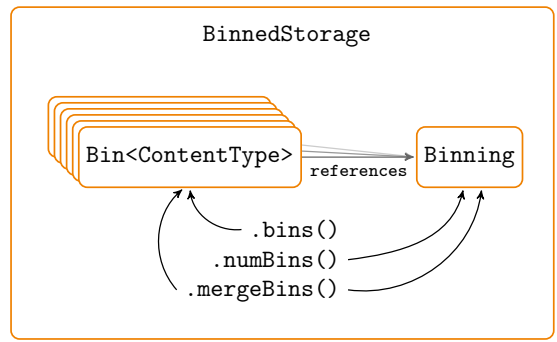

The types introduced so far permit histogramming to be implemented
using separate bin-location and bin-content facilities, but the user
would need to manually maintain consistency between them. To ensure
this automatically, a generic |BinnedStorage| class is introduced,
illustrated in Figure~\ref{fig:storage-diagram}. This type contains
both a |Binning|, constructed from a \cxx parameter pack of the edge
types defining the fill space, and stored bin-content as a contiguous
array of |Bin|-wrapped content types. The |BinnedStorage|
%uses the previously discussed |Binning| class to define and
manages its fill space, providing access to individual bins via a
|bin| method which accepts the global bin index or the array of local
indices, a |binAt| method for locating the bin via a set of \fillcoord
coordinate values, and a |bins| method which returns a wrapping vector
type that can elide overflow and masked bins depending on optional
arguments. The |BinnedStorage| also ensures consistency when
redefining binnings, e.g.~via the |mergeBins| method which
simultaneously eliminates edges across a range of bins, and merges the
corresponding content objects.

This structure, however, is still too generic for user-facing standard
histograms. In order to achieve a live, update-able specialisation,
an additional |FillableStorage| inheritance layer is introduced,
providing the means to update its internal statistics using the
\cxx adaptor pattern. A \emph{fill adaptor} type can be passed in by
the user, allowing customisation of the \emph{fill} concept depending
on the specifics of the bin-content type in question. Whenever the
user calls |fill| on the object, passing a set of coordinates and an
optional fill-weight and fill-fraction, the |FillableStorage| uses its
contained |Binning| object to identify the targeted bin,
%that corresponds to the set of fill-space coordindates provided.
then passes its global index to the fill-adaptor object, which handles
the specialised updating of the bin's content. At the end of the
|fill| call, the global bin index is returned. If any of the
coordinates is |nan|, it is in general not possible to identify the
corresponding bin and so a |-1| is returned, and the |nan|-fill
recorded.  If a bin can be identified but is masked, the bin
statistics will still be updated and the position returned. This gives
the user more flexibility to change their mind about unmasking or
merging bins at a later time.  The |FillableStorage| also adds the
fill-dimensionality as an additional template parameter, allowing in
general the |fill| method to accept more coordinates than the binning
dimensionality, e.g.~enabling a unified implementation of histogram
and generalised-profile live binned data-types.

\subsection{Standard histograms and profiles}

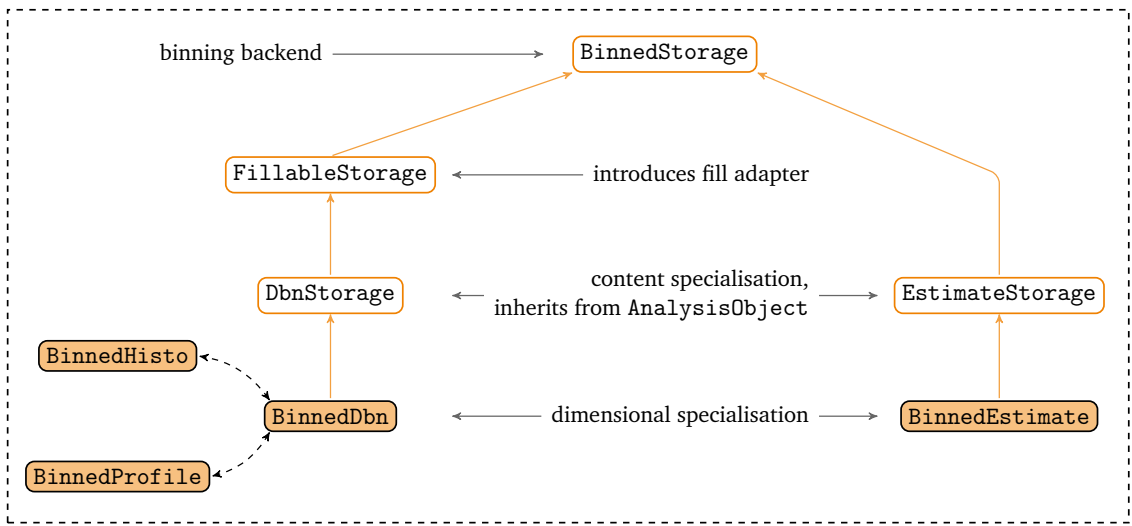
\begin{figure}%[htb!]
 \input{dev-diagram}
 \caption{Internal binned-type class inheritance structure in
   \yoda[2]. The diagram illustrates how several layers of inheritance
   and template-specialisation (orange arrows) are used to build
   parallel live and inert data-types, with higher layers more generic
   and lower ones more specialised and user-facing for standard
   statistical operations, in particular the specialised aliases
   connected via dashed lines on the bottom-left.}
 \label{fig:dev-diagram}
\end{figure}

Further stages of specialisation, illustrated in
Figure~\ref{fig:dev-diagram}, are used to reduce this highly
general picture of binned and fillable data-storage to the familiar
statistical types discussed in Section~\ref{sec:stats}, presenting a
user-facing view of the most commonly used statistical machinery. The
desired live and inert types are respectively implemented as a
|FillableStorage| containing |Dbn| objects and a |BinnedStorage|
containing |Estimate| objects. Again inheritance and template
specialisations are used to provide features and reduce code
duplication, with derived |DbnStorage| and |EstimateStorage| types
adding whole-object facilities such as integrals over the live |Dbns|,
or area-under-curve or estimate-averaging facilities on the
collections of inert binned estimates. At this point, with the content
types fully specified, the data-objects also acquire an inheritance
relationship from the virtual |AnalysisObject| base class which
provides a metadata storage utility and integration with the I/O
persistency systems to be described in Section~\ref{sec:io}.

The very final step is to provide some user-friendliness refinements
for the most familiar types: the |BinnedEstimate| and |BinnedDbn|
inheritance layers employ the CRTP to mix in axis-specific method
names for the first three binning dimensions, e.g.~|xEdges| or
|sumWY2|.  %% The full set of layers for both live and inert types is
%% illustrated in Figure~\ref{fig:user-diagram}. %% for the case of live
%% objects and would
%% look the same for inert objects, except that the |FillableStorage|
%% layer is not used in the inert case of course.

\begin{figure}%[htb!]
 \input{user-diagram}
 \caption{User-facing class inheritance structure in \yoda[2]. The
   orange lines show that all these types inherit from the persistency
   and metadata features of \texttt{AnalysisObject}, and the black
   arrows indicate the direction of live-to-inert
   type-reductions. Direct type reductions to \texttt{ScatterND} from
   \texttt{Counter} and \texttt{BinnedDbn} are also allowed, bypassing
   the intermediate \texttt{Estimate}s.}
 \label{fig:user-diagram}
\end{figure}
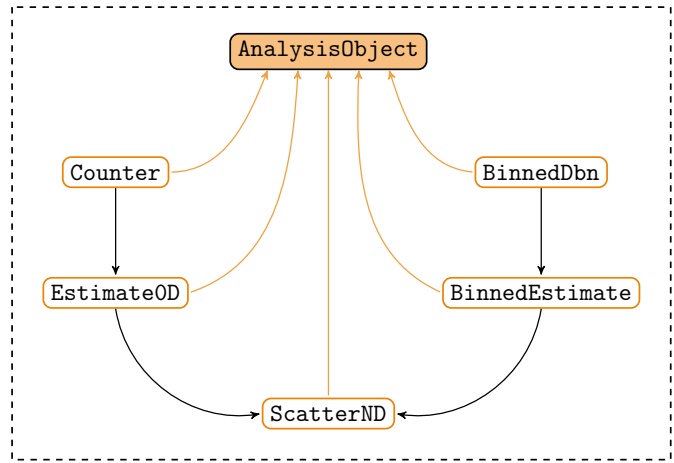

Various aliases are defined to enable convenient shorthands, with the
whole user-facing type family shown in Figure~\ref{fig:user-diagram}:
\begin{itemize}
  \item |BinnedHisto| takes a parameter pack of edge types and is an
    alias for a |BinnedDbn| with its fill dimensionality given by the
    number of elements in the parameter pack.
  \item |BinnedProfile| takes a parameter pack of edge types and is an
    alias for a |BinnedDbn| with its fill dimensionality given by the
    number of elements in the parameter pack plus one, for the
    unbinned axis.
  \item |HistoND| and |ProfileND| are both templated on a positive
    integer that is internally translated into a corresponding number
    of |double|s, as a shorthand for a histogram or profile with only
    continuous axes, respectively.
  \item For the first three dimensions, more legible aliases
    |Histo1D|, |Profile2D| etc.  are also defined, largely for
    familiarity with the previous \yoda[1] implementations.
\end{itemize}

The full suite of natively supported user-facing types inheriting from
|AnalysisObject| is illustrated in Figure~\ref{fig:user-diagram}.  The
|BinnedDbn| and |BinnedEstimate| are accompanied by unbinned |Counter|
and |Estimate0D| objects which have no fill-space and hence correspond
to the unbinned summary statistics from the start of
Section~\ref{sec:stats}; these are implemented separately due to their
simplicity.

Timing performance tests were made using a similar methodology to
\boosth (BH), using 1M samplings per dimension from a unit Gaussian
located at $\mu = 1$. These were filled into histograms of
dimensionality~1 to~5, each axis containing 20 bins either uniform in
$[-2,2]$ or logarithmic from $[0.02, 2]$, ensuring the non-linear
search and overflow features were tested. Linear scaling of fill time
was observed in single-threaded running on a 4.0~GHz CPU, from
0.19~(0.21)~secs in 1D to 0.94~(0.99)~secs in 5D for 1M fills into
linear (logarithmic) binnings respectively. While histogramming is
rarely a performance bottleneck and this speed is sufficient for
analysis use-cases, it is slower than BH to the extent comparisons can
be made given the difference between full-moments histogramming in
\yoda[2] vs integer-count fills in the BH benchmarking. This remains
an area for possible performance optimisation.

\subsection{Type and dimensionality reductions}
At the end of a run, the live types can be reduced to their inert
equivalents and can be subjected to further post-processing, e.g.~in
order to combine systematic variation runs. Inert types are also the
natural output from object-combining operations such as histogram
division or efficiency computation (differing by uncertainty
treatment).  The inert types lend themselves naturally to representing
the experimental cross-section measurements available on
\hepdata~\cite{Maguire:2017ypu}, and the built-in |covarianceMatrix|
method of the |BinnedEstimate| class can be used to translate these
error breakdowns into a covariance.  When reducing a live type to an
inert type, masked bins are skipped, thereby retaining the default
|nan| of the corresponding |Estimate| objects contained within the
|BinnedEstimate|, which can then be interpreted as a visual gap by
plotting scripts.

Both live and inert types can also be reduced to the |ScatterND| class
which contains an array of |PointND| objects -- a set of coordinates
and negative/positive uncertainty pairs in $N$ dimensions, which lends
itself directly to the concept of a marker on the canvas.  When
converting an |Estimate| to a |PointND|, the total uncertainty is
transferred and the error breakdown is lost. Bin-focus information is
lost from live objects if converting to a scatter via an estimate, as
the inert binning-content has no such concept, but can be preserved if
reducing directly from a live type to a point-based scatter.

All inert types, including scatters, can also be handed functors such
as \cxx lambda functions in order to apply arbitrary transformations
on the inert statistical distribution. These will operate on the
underlying |PointND| or |Estimate| objects, with consistent non-linear
propagation to both values and errors, with appropriate distinct
treatment of correlated and uncorrelated error-source in the latter
case.

It is also possible to switch between different types of |BinnedDbn|,
typically by reducing either the fill- or binning dimensionality in
the process. For instance, a histogram with $N_\mathrm{ax}$ binned
axes can be created from a profile with the same axes by simply
dropping the unbinned axis using the |mkHisto| method.
Marginalisations across a given binned axis can be achieved using the
|mkMarginalHisto| and |mkMarginalProfile| methods, which also reduce
the dimensionality of the contained |Dbn| type, along with the binned
dimensionality. In addition, it is possible to \defn{slice} a
|BinnedStorage| along one of the binned axes into a vector of
|BinnedStorage| objects that are have their binning dimensionality
reduced by one unit. The resulting vector will have one element for
each bin along the axis that is being sliced over by calling
|mkHistos|, |mkProfiles| or |mkEstimates|.

Finally, note that because the underlying |BinnedStorage| can contain
arbitrary types, it is even possible to construct \emph{histogram
groups}, implemented as a |BinnedStorage| containing more
|BinnedStorage| objects. This can be useful when trying to represent a
2D histogram as a group of 1D histograms connected via an implicit
group axis.

%\item merging a visible and masked bin will result in an unmasked bin

\subsection{\python interface layer}

A \python interface is provided through \cython bindings to the
underlying \cxx implementation.  The \cython source files are
automatically generated from a script at build time up to a given
number of dimensions and a set of axis edge types.  Note that full
support of arbitrary dimensions like in \cxx due to its compile-time
polymorphism is currently not possible in \python due to its
dynamically-typed nature.  Whenever lists are returned by a method, an
automatic conversion to \numpy arrays is attempted first with a
fall-back to the default \python list type in case \numpy cannot be
found. ``Pythonic'' customisation of the API has been avoided, as
the experience of \yoda[1] was that distinct APIs for each language
was more confusing than helpful.

%% file: binning-diagram.tex
\begin{center}
\scalebox{0.8}{
%% \tikzstyle{feed}=[shorten >=1pt,>=stealth',semithick]
%% \tikzstyle{edgeLabel} = [pos=0.5, text centered] %, font={\sffamily\small}]
\begin{tikzpicture}[bend angle=45, thick]
\begin{scope}[xshift=0cm,yshift=0cm]

\fill[lightgray!50] (-3, -3) rectangle (3,3);
\fill[lightgray] (-0.5, 0.5) rectangle (2.5, 2.0);
  
\draw[dashed,<-] (-4.0, 3.0) node[left]{$-\infty$} -- (-3.0, 3.0); \draw[solid] (-3.0, 3.0) -- (3.0, 3.0); \draw[dashed,->] (3.0,3.0) -- (4.0, 3.0) node[right]{$+\infty$};
\draw[dashed,<-] (-4.0, 2.0) -- (-3.0, 2.0); \draw[solid] (-3.0, 2.0) -- (3.0, 2.0); \draw[dashed,->] (3.0, 2.0) -- (4.0, 2.0);
\draw[dashed,<-] (-4.0, 0.5) -- (-3.0, 0.5); \draw[solid] (-3.0, 0.5) -- (3.0, 0.5); \draw[dashed,->] (3.0, 0.5) -- (4.0, 0.5);
\draw[dashed,<-] (-4.0,-2.5) -- (-3.0,-2.5); \draw[solid] (-3.0,-2.5) -- (3.0,-2.5); \draw[dashed,->] (3.0,-2.5) -- (4.0,-2.5);
\draw[dashed,<-] (-4.0,-3.0) -- (-3.0,-3.0); \draw[solid] (-3.0,-3.0) -- (3.0,-3.0); \draw[dashed,->] (3.0,-3.0) -- (4.0,-3.0);

\draw[dashed,<-] (-3.0, -4.0) node[below]{$-\infty$} -- (-3.0, -3.0); \draw[solid] (-3.0, -3.0) -- (-3.0, 3.0); \draw[dashed,->] (-3.0,3.0) -- (-3.0,4.0) node[above]{$+\infty$};
\draw[dashed,<-] (-1.0,-4.0) -- (-1.0, -3.0); \draw[solid] (-1.0, -3.0) -- (-1.0, 3.0); \draw[dashed,->] (-1.0,3.0) -- (-1.0, 4.0);
\draw[dashed,<-] (-0.5,-4.0) -- (-0.5, -3.0); \draw[solid] (-0.5, -3.0) -- (-0.5, 3.0); \draw[dashed,->] (-0.5,3.0) -- (-0.5, 4.0);
\draw[dashed,<-] ( 2.5,-4.0) -- ( 2.5, -3.0); \draw[solid] ( 2.5, -3.0) -- ( 2.5, 3.0); \draw[dashed,->] ( 2.5,3.0) -- ( 2.5, 4.0);
\draw[dashed,<-] ( 3.0,-4.0) -- ( 3.0, -3.0); \draw[solid] ( 3.0, -3.0) -- ( 3.0, 3.0); \draw[dashed,->] ( 3.0,3.0) -- ( 3.0, 4.0);

\node at (-3.7,-3.7) {$(0,0)$};
\node at (-2.0,-3.7) {$(1,0)$};
\node at (-3.7,-1.0) {$(0,2)$};
\node at (+1.0,-1.0) {$(3,2)$};
\node at (+1.0,+1.5) {$(3,3)$};
\node at (+1.0,+1.0) {masked};

\end{scope}
\end{tikzpicture}}
\end{center}

%% file: storage-diagram.tex
\begin{center}
\scalebox{0.8}{
\tikzstyle{feed}=[shorten >=1pt,>=stealth',semithick]
\tikzstyle{edgeLabel} = [pos=0.5, text centered] %, font={\sffamily\small}]
\begin{tikzpicture}[bend angle = 45]

\begin{scope}[xshift=0cm,yshift=0cm]

\node (Storage) [text centered] at (3.0,4.0) {\texttt{BinnedStorage}};

\foreach [evaluate=\p as \x using 0.8+\p*0.1, evaluate=\p as \y using 2.6-\p*0.1] \p in {1,...,6} {

\node (Bin) [rectangle, fill = white, rounded corners, minimum width=1.5cm, minimum height=1.0cm,text centered, thick, draw=irn-bru,
] at (\x,\y) {\texttt{Bin<ContentType>}};

}

\node (Binning) [rectangle, rounded corners, minimum width=1.5cm, minimum height=1.0cm,align=center,text centered, thick, draw=irn-bru,
] at (6.0,2.0) {\texttt{Binning}};

\node (M-bins) [left,
] at (4.0,0.8) {\texttt{.bins()}};

\node (M-numBins) [left,
] at (4.0,0.3) {\texttt{.numBins()}};

\node (M-merge) [left,
] at (4.0,-0.2) {\texttt{.mergeBins()}};

\draw[thick,rectangle,rounded corners,draw=irn-bru] ($(Bin.south west)+(-1.1,+3.0)$) rectangle ($(Binning.south east)+(0.5,-2.5)$);

\draw[solid]

(Bin.east) edge [feed,->,yshift=+0.2cm,black!20] (Binning.west)

(Bin.east) edge [feed,->,yshift=+0.1cm,black!40] (Binning.west);

\draw [feed,->,black!60] (Bin.east) -- (Binning.west) node[midway,below,black,font=\scriptsize] {\texttt{references}};

\draw[solid]

(M-bins.west) edge [feed,->,bend left] (Bin.south)

%(M-numBins.east) edge [feed,->,bend right,out=-22.5] ($(Binning.south)+(-0.1,0.0)$)
%(M-numBins.east) edge [feed,->,bend right] ($(Binning.south)+(-0.1,0.0)$)
(M-numBins.east) edge [feed,->,bend right,out=-25,in=-135] ($(Binning.south)+(-0.1,0.0)$)

(M-merge.west) edge [feed,->,bend left] ($(Bin.south)+(-0.1,0.0)$)

(M-merge.east) edge [feed,->,bend right,out=-45] ($(Binning.south)+(+0.2,0.0)$)

;

\end{scope}
\end{tikzpicture}}
\end{center}

%% file: dev-diagram.tex
% All histograms can be instantiated through this alias.
%
%               BinnedStorage                : Introduces binning backend.
%                     |
%              FillableStorage               : Introduces FillAdapterT
%                     |
%                 DbnStorage                 : Hooks up with AnalysisObject
%                     /\
%                    /  \
%    BinnedDbn<1>___/    \___ BinnedDbn<2>   : Introduces dimension specific
%          \                     /           : utility aliases
%           \_______     _______/            : (xMin(), yMax(), etc.)
%                   \   /
%                    \ /
%                     |
%         BinnedHisto/BinnedProfile          : Convenience alias
%
% Since objects with continuous axes are by far the most commonly used type
% in practice, we define convenient short-hand aliases HistoND/ProfileND for
% with only continuous axes, along with the familar types Histo1D, Profile2D, etc.

\begin{center}
\scalebox{0.8}{
\tikzstyle{feed}=[shorten >=1pt,>=stealth',semithick]
\tikzstyle{edgeLabel} = [pos=0.5, text centered] %, font={\sffamily\small}]
\begin{tikzpicture}[bend angle = 45]

\begin{scope}[xshift=0cm,yshift=0cm]

\node (BS) [rectangle, rounded corners, minimum width=1.5cm, minimum height=0.5cm,text centered, thick, draw=irn-bru,
] at (9.5,6.5) {\texttt{BinnedStorage}};

\node (FS) [rectangle, rounded corners, minimum width=1.5cm, minimum height=0.5cm,text centered, thick, draw=irn-bru,
] at (4.0,4.5) {\texttt{FillableStorage}};

\node (DS) [rectangle, rounded corners, minimum width=1.5cm, minimum height=0.5cm,text centered, thick, draw=irn-bru,
] at (4.0,2.5) {\texttt{DbnStorage}};

\node (BND) [rectangle, rounded corners, minimum width=1.5cm, minimum height=0.5cm,text centered, thick, draw=black,
fill=irn-bru!50] at (4.0,0.5) {\texttt{BinnedDbn}};

\node (BH) [rectangle, rounded corners, minimum width=1.5cm, minimum height=0.5cm,text centered, thick, draw=black,
fill=irn-bru!50] at (0.5,1.5) {\texttt{BinnedHisto}};

\node (BP) [rectangle, rounded corners, minimum width=1.5cm, minimum height=0.5cm,text centered, thick, draw=black,
fill=irn-bru!50] at (0.5,-0.5) {\texttt{BinnedProfile}};

\node (FS2) at (15.0,4.5) {};

\node (DS2) [rectangle, rounded corners, minimum width=1.5cm, minimum height=0.5cm,text centered, thick, draw=irn-bru,
] at (15.0,2.5) {\texttt{EstimateStorage}};

\node (BND2) [rectangle, rounded corners, minimum width=1.5cm, minimum height=0.5cm,text centered, thick, draw=black,
fill=irn-bru!50] at (15.0,0.5) {\texttt{BinnedEstimate}};

\node (LBS) [left,
] at (4.0,6.5) {binning backend};

\node (LFS) [left,
] at (12.0,4.5) {introduces fill adapter};

\node (LDS) [left, align=right,
] at (12.0,2.5) {content specialisation,\\ inherits from \texttt{AnalysisObject}};

\node (LBND) [left,
] at (12.0,0.5) {dimensional specialisation};

\draw[thick,dashed] ($(BH.south west)+(-0.5,-2.5)$) rectangle ($(BND2.south east)+(0.5,7.0)$);

\draw[solid]

(BS.south west) edge [feed,<-,irn-bru!75] (FS.north)

(FS.south) edge [feed,<-,irn-bru!75] (DS.north)

(DS.south) edge [feed,<-,irn-bru!75] (BND.north)

(DS2.south) edge [feed,<-,irn-bru!75] (BND2.north)

($(BND.north)+(-1.0,0.0)$) edge [feed,<->,dashed,bend right,out=-22.5,in=-157.5] (BH.east)

($(BND.south)+(-1.0,0.0)$) edge [feed,<->,dashed,bend left,out=22.5,in=157.5] (BP.east)

($(FS)+(2.0,0.0)$) edge [feed,<-,draw=black!60] (LFS.west)

($(DS)+(2.0,0.0)$) edge [feed,<-,draw=black!60] (LDS.west)

($(BND)+(2.0,0.0)$) edge [feed,<-,draw=black!60] (LBND.west)

($(BS)+(-2.0,0.0)$) edge [feed,<-,draw=black!60] (LBS.east)

($(DS2)+(-2.0,0.0)$) edge [feed,<-,draw=black!60] (LDS.east)

($(BND2)+(-2.0,0.0)$) edge [feed,<-,draw=black!60] (LBND.east)

;

\draw[feed,rounded corners,<-,irn-bru!75] (BS.south east) -- (FS2.center) -- (DS2.north);

\end{scope}
\end{tikzpicture}}
\end{center}

%% file: user-diagram.tex
%----------------------
%|   AnalysisObject   |
%---------------------
%
%Counter       BinnedDbn
%   |              |
%   |              |
%Estimate0D  BinnedEstimate
%   \              /
%     \           /
%       ScatterND

\begin{center}
\scalebox{0.8}{
\tikzstyle{feed}=[shorten >=1pt,>=stealth',semithick]
\tikzstyle{edgeLabel} = [pos=0.5, text centered] %, font={\sffamily\small}]
\begin{tikzpicture}[bend angle = 45]

\begin{scope}[xshift=0cm,yshift=0cm]

\node (AO) [rectangle, rounded corners, minimum width=1.5cm, minimum height=0.5cm,text centered, thick, draw=black,
fill=irn-bru!50] at (4.0,5.5) {\texttt{AnalysisObject}};

\node (C) [rectangle, rounded corners, minimum width=1.5cm, minimum height=0.5cm,text centered, draw=irn-bru, thick,
] at (0.5,3.5) {\texttt{Counter}};

\node (BD) [rectangle, rounded corners, minimum width=1.5cm, minimum height=0.5cm,text centered, draw=irn-bru, thick,
] at (7.5,3.5) {\texttt{BinnedDbn}};

\node (E0D) [rectangle, rounded corners, minimum width=1.5cm, minimum height=0.5cm,text centered, draw=irn-bru, thick,
] at (0.5,1.5) {\texttt{Estimate0D}};

\node (BE) [rectangle, rounded corners, minimum width=1.5cm, minimum height=0.5cm,text centered, draw=irn-bru, thick,
] at (7.5,1.5) {\texttt{BinnedEstimate}};

\node (SND) [rectangle, rounded corners, minimum width=1.5cm, minimum height=0.5cm,text centered, draw=irn-bru, thick,
] at (4.0,-0.5) {\texttt{ScatterND}};

\draw[thick,dashed] ($(E0D.south west)+(-0.5,5.0)$) rectangle ($(BE.south east)+(0.5,-2.5)$);

\draw[solid]

($(AO.south)+(-1.0,0.0)$) edge [feed,<-,bend left,out=22.5,irn-bru!75] (C.east)

($(AO.south)+(1.0,0.0)$) edge [feed,<-,bend right,out=-22.5,irn-bru!75] (BD.west)

($(AO.south)+(-0.5,0.0)$) edge [feed,<-,bend left,out=22.5,irn-bru!75] (E0D.east)

($(AO.south)+(0.5,0.0)$) edge [feed,<-,bend right,out=-22.5,irn-bru!75] (BE.west)

(AO.south) edge [feed,<-,irn-bru!75] (SND.north)

(C.south) edge [feed,->] (E0D.north)

(BD.south) edge [feed,->] (BE.north)

(E0D.south) edge [feed,->,bend right] (SND.west)

(BE.south) edge [feed,->,bend left] (SND.east);

\end{scope}
\end{tikzpicture}}
\end{center}

%% file: sec-formats.tex
\section{Data exchange and persistency}
\label{sec:io}
\label{sec:formats}
\label{sec:persistency}

The |AnalysisObject| type is the key link between in-memory \yoda
data-objects and being able to read and write persistent
representations of those objects.  There is no requirement to do so:
it is perfectly valid to work with in-memory \yoda objects only, and
unlike e.g.~\ROOT there is no mandatory binding of histograms and
other data objects to a global I/O registry, but of course in practice
much of the time users do need to be able to write out their data one
way or another.

\subsection{Attributes and I/O}

The |AnalysisObject| base class provides a generic ``attributes''
system for storing arbitrary metadata. The fundamental attribute
storage is (for now) string-based and hence not designed for
high-performance storage of additional numeric data, but does provide
a very useful way to store information such as normalization-scaling
history, original data types (in the case of live-to-inert type
reductions), plotting directives such as axis labels or legend titles,
and a unique Unix-like absolute path for each object. The \python
interface uses the \sw{PyYAML} package~\cite{pyyaml}, or the |ast|
standard-library module to automatically convert the types of
e.g.~numerical lists stored as attributes.

The object paths are used by format-specific |Writer| classes to
identify the persisted objects, and as the keys of a map returned by a
singleton |Reader| class. A type register is used to help bridge the
gap between the \cxx compile-time templates and the desire to
dynamically declare object types.  The most frequent object types and
set of axis edge types are preregistered when the |Reader| singleton
is first instantiated and can then be augmented at run time with more
complex types using the |registerType<type>()| method.

I/O operations can either be performed by explicitly instantiating
these I/O manager objects, or implicitly via unbound |write| and
|read| functions. At present, a structured-text format with support
for \sw{gzip} I/O is the main \yoda persistency scheme, extended for
version~2 to store bin-edges in a non-duplicating way for improved
efficiency; other supported formats include a more readily
human-readable representation for inspection and a ``flat'' format for
uniform inspection of data-objects as inert scatter types. Only the
structured-text format is currently supported by the |Reader|
interface, and support for the defunct AIDA XML data format has been
dropped; an HDF5 read/write format is planned~\cite{Folk1999HDF5}.

Inspection and manipulation of \yoda data files is aided by a suite of
command-line tools including |yodals| for listing (in several levels
of detail) file contents, and converter scripts e.g.~|yoda2root| and
|root2yoda|, and a |yoda2yoda| utility mainly useful for allowing
users to filter (both positively and negatively) which objects are
copied to the destination file.

In this supporting tool-suite and associated tools, we have used
natural extensions of the Unix-path concept to index bins or points
within the objects as well via suffix operators: |/my/histo| refers to
a histogram, |/my/histo#3| to its bin with global-index |3|,
|/my/histo2#3,4| to a bin in a 2D histogram via local indices,
|/my/histo2@1.5,10.4| for coordinate-based indexing cf.~the |binAt|
method, and in general ranges of bins via e.g.~|/my/histo#3@25|. A
convention also in use is of square-bracketed path suffixes (before
any bin specification) to indicate systematic variations on a nominal
data object, e.g.~|/my/histo[SCALE_VARIATION]|. We argue that this
unified approach to histogram and bin identifiers is a generally
useful concept for binned-data analysis tools and welcome wider uptake
or discussion.

\subsection{Serialisation and MPI syncing}

The |AnalysisObject| additionally provides virtual methods names
|serializeContent| and |deserializeContent|, to allow for translation
of its derived object types to and from an |std::vector<double>|. This
format lends itself better to data-gathering operations as part of MPI
collective communications across different process ranks.  Equivalent
|serializeMeta| and |deserializeMeta| methods exist to achieve a
similar (de-)construction of the object's attribute metadata to and
from an |std::vector<std::string>|, which in case of estimates would
include the error-source labels. Serialization is primarily aimed at
high-performance numerical use-cases, with a normal expectation that
attributes will match between the objects being synchronised/gathered
across ranks, hence the relative inefficiency of string-based metadata
serialization need not be encountered in most applications.

%% file: sec-plot.tex
\section{Plotting and visualisation}
\label{sec:plot}
\label{sec:plotting}

\yoda offers full functionality for plotting and visualising its
different datatypes.  The release of \yoda[2] comes with a \python API
plotting back-end, providing an interface with the dominant modern
data-visualisation toolkit (\mpl) and computational libraries
(e.g.~\numpy). The plotting API is designed to be quick and to produce
highly customisable plots, ensuring these features can easily be
interfaced to external software. For users wishing to write their own
plotting code, the Python interface provides single-call access to
the data typically needed by \mpl functions.

The \yoda plotting module supports various customisations of plot
features, such as the generation of ratio plots, switching between
linear or logarithmic axes, the customisation of plot labels and
more. This interface is designed to facilitate the interaction of
\yoda objects and \mpl, and to provide the user with out-of-the-box
comparison plots in a pre-wrapped common style. As detailed below,
further customisations from this starting point are possible through
the \mpl interface.

\begin{figure}%[htb!]
 \includegraphics[width=0.48\textwidth]{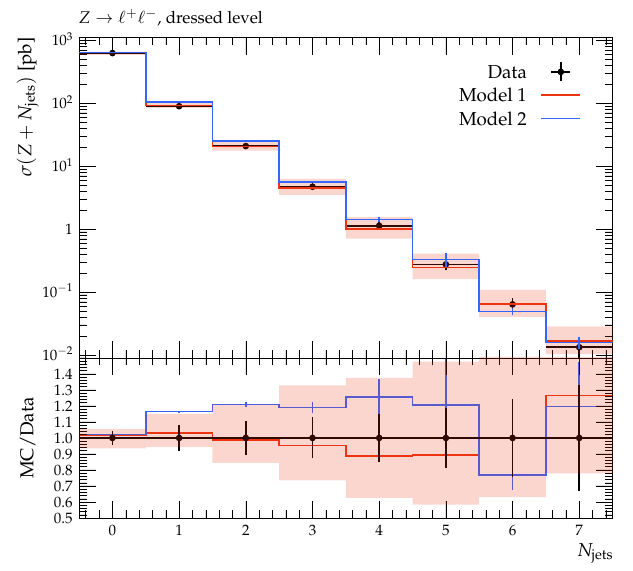}\quad
 \includegraphics[width=0.48\textwidth]{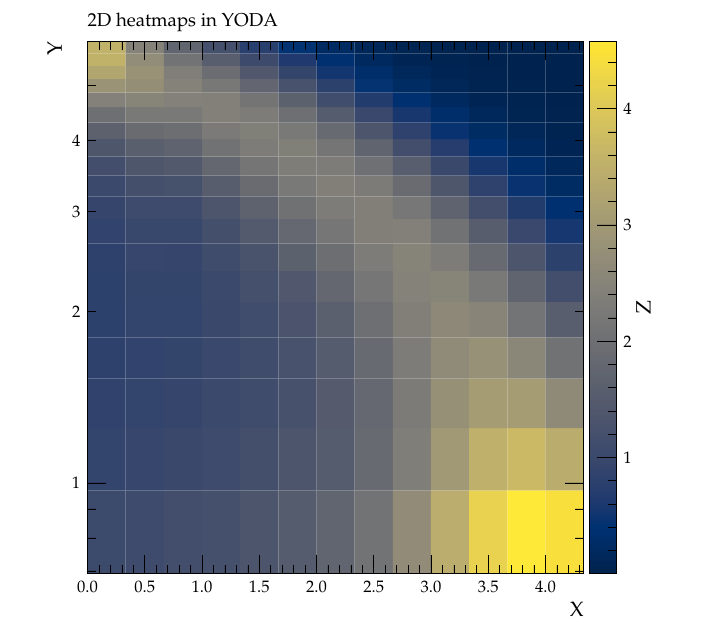}
 \caption{Example 1D and 2D plot outputs from the \yoda[2] plotting
   interface, illustrating a mix of uniform and irregular bin sizes,
   and automatic use of ratio plotting for 1D data/prediction comparisons.}
 \label{fig:yplots}
\end{figure}

The simplest way to plot \yoda objects is with the |yodaplot|
executable, which takes any number of \yoda files as input. For 1D
histograms, it will identify the matching object paths from the
different \yoda files and overlay them. This ultimately generates a
single plot for each object path, i.e.~the |Scatter2D| objects, where
entries from multiple \yoda files are shown in a comparison plot.  By
default an additional ratio panel is drawn, illustrating the agreement
relative to the \yoda file that was specified first on the
command-line.  The ratio panel can be disabled via a command-line
flag.  Two-dimensional histograms -- with two independent variables
and one dependent variable -- are drawn on to separate canvases, so
that 2D heatmaps are created for each object path (corresponding
to~|Scatter3D| objects) for every input \yoda file. Example one- and
two-dimensional plots are shown in Figure~\ref{fig:yplots}.

% \TODO{Mention 2D plotting options: did we make it possible to do
%   e.g. a heatmap with contours overlaid from other file(s), or is that
%   something where API use is needed? I think worth saying a bit about
%   the possibility of complex things, i.e.~at some level this interface
%   is designed to ease the interaction of \yoda objects and \sw{mpl},
%   and to provide pre-wrapped common comparison-plot styles, but API
%   users can always fall back to custom \sw{mpl} operations on the
%   resulting objects.}

A distinctive plotting feature in \yoda[2] is that in addition to the
graphical outputs, for example in PDF or PNG format, the API generates
a set of executable \python scripts via a ``script generator'' system.
This is configured via a plot-specification dictionary that connects
\yoda data-objects to graphical representation preferences; e.g., for
an analysis object |ao| read from file |fname|, a standalone \mpl script
can be generated using
\begin{lstlisting}[language=Python]
 plotspec = { "histograms" : { fname : { "nominal" : ao } } } 
 script_generator.process(plotspec, "histo_name")
\end{lstlisting}
where |script_generator| is an instance of the script-generation
manager class.

These generated scripts contain low-level plotting-calls to the \mpl API
and can be executed to regenerate the graphical output -- indeed, the
graphical output of the commands is accomplished through execution of
these scripts, with some optimisations using parallel processing and a
shared single import of the \mpl library for efficiency.  For
maintainability, the plotting data is isolated in a separate |.py|
file from the presentation commands, again providing a degree of
separation between style and data.

The |yodaplot| script itself is a pedagogical example of the utilities
offered by the Python API. After loading the relevant module at the
start of the script, a nested loop over the input \yoda files and
object paths generates the Python plotting scripts through the API.
In the last step, |yodaplot| executes the plotting scripts (with
support for parallel processing) to generate the graphical outputs.
Other software tools wrapping around the \yoda plotting API include
\rivet and \contur. In these examples, a pre-processing step collates
information on the plotting style and data in a Python dictionary,
which is then processed by the \yoda |script_generator| function to
generate executable Python scripts for plotting.

Not only does the text-based intermediate format offer a
platform-independent and future-proof way to regenerate and to finesse
graphical outputs, e.g.~as required by iterative review processes, but
the scripts are also designed to be transparent and easy to understand
for the user. This design was informed by positive experience with a
text-based plot-summary data format in the Rivet package's legacy
\TeX-based plotting system, which has now been replaced by this new
implementation. As the underlying \mpl objects -- figures, axes,
lines, etc. -- are accessible in the generated scripts, arbitrary
customisations and refinements are possible.

%% file: sec-concl.tex
\section{Conclusions and outlook}
\label{sec:concl}
\label{sec:conclusion}

We have introduced the \yoda[2] library for consistent,
high-dimensional, and flexible histogramming, expressed in type-safe
modern \cxx. The expression of dimensionality and data-type
relationships enables aggressive compile-time performance
optimisation, while significant effort has been expended to ensure
that the majority of technical complexity is hidden in normal usage.

A strong emphasis is placed on separation of the binning concept from
aggregation of statistical moments, making possible both storage of
arbitrary object types in efficient partitions of independent
variables, and exact conversion between different projections of
second-order statistical cumulants. Live and inert data-types are
strictly separated, with one-directional conversion mechanisms and
a set of simple |Scatter| data-types used as the basis for a modern
\python-based plotting interface.

Persistency is currently supported to a text-based (and gzipped)
custom format, with work underway to provide a more performant HDF5
format. This latter will complement current features for over-the-wire
serialisation in MPI messages, making \yoda[2] ideally suited for
large-scale operation on high-performance computing clusters.

The \yoda[2] library is already in production use with the Rivet MC
analysis tool for particle physics, but is of completely general
design and suited to any application in need of performant and
low-dependency binned statistics or data-storage. Its programmatic
usage from \cxx and \python is complemented also by a set of
command-line tools for dataset inspection, manipulation and
combination.

Several future developments are planned, in addition to the already
mentioned HDF5 data-format. A common method
%in particle physics
for encoding of systematic uncertainties is to propagate not just a
single weight $w_n$ for each fill, but a family of them,
$\{w_n^{(i)}\}$, where the $i$ indexes a set of \defn{weight streams}
that coherently reflect the change of the $n$'th fill's probability
under different model assumptions. At present such uncertainties must
be encoded through extended suffixes in the \yoda |AnalysisObject|
paths, but a more elegant scheme would generalise the |DbnStorage|
classes to be implicitly multi-weighted, without the overheads of
maintaining parallel copies of entire histograms (and performing
repeated binning lookups). Another natural area for extension is to
support an unbinned multidimensional data type, analogous to the
``ntuple'' types commonly used in particle physics and many other
data-science applications. Finally, we note that the template
meta-programming techniques used to implement \yoda[2] are
increasingly embraced and enhanced within the \cxx language, and we
anticipate further growth of the power, expressiveness, and
performance of the \yoda system along with future language evolutions.